# Optical NEP in Hot-Electron Nanobolometers

Boris S. Karasik[1*] and Robin Cantor[2]

[1]Jet Propulsion Laboratory, California Institute of Technology, Pasadena, California, CA 91109, USA

[2]STAR Cryoelectronics, 25-A Bisbee Court, Santa Fe, NM 87508, USA

*Contact: boris.s.karasik@jpl.nasa.gov, phone +1-818-393 4438

*Abstract*—For the first time, we have measured the optical noise equivalent power (NEP) in titanium (Ti) superconducting hot-electron nanobolometers (nano-HEBs). The bolometers were 2µm×1µm×20nm and 1µm×1µm×20nm planar antenna-coupled devices. The measurements were done at $\lambda$ = 460 µm using a cryogenic black body radiation source delivering optical power from a fraction of a femtowatt to a few 100s of femtowatts. A record low NEP = $3\times10^{-19}$ W/Hz$^{1/2}$ at 50 mK has been achieved. This sensitivity meets the requirements for SAFARI instrument on the SPICA telescope. The ways for further improvement of the nano-HEB detector sensitivity are discussed.

## I. Introduction

Recently, an interest to ultrasensitive submillimeter/FIR detectors has been driven due to the astrophysics community and space agencies' plans to launch telescopes with cryogenically cooled primary mirrors where the thermal emission from the telescope would be largely eliminated. Several concepts of such telescopes (SAFIR [1-2], SPECS [2], CALISTO [3], FIRI [4], SPICA [5]) have been proposed and studied. Moderate resolution spectroscopy ($\nu/\delta\nu \sim 1000$) would be the most demanding application requiring the detector NEP to be less than $10^{-19}$ W/Hz$^{1/2}$ in the most of the submillimeter/FIR spectral range in order for the photon shot noise to dominate the detector noise. The SPICA mission led by the Japanese space agency [6] is currently seen as the most feasible opportunity to realize such a sensitive astronomical platform. There are two instruments considered for far-IR spectroscopy on SPICA. An ESA led SAFARI instrument would use a Fourier Transform Spectrometer in the 30-210 µm wavelength range [7]. The detectors for SAFARI should have an NEP = $3\times10^{-19}$ W/Hz$^{1/2}$. Another instrument (BLISS) [8], which is under study in the US, is a grating spectrometer operating in the 35-433 µm range. The detector sensitivity goal here (NEP = $3\times10^{-20}$ W/Hz$^{1/2}$) is non-precedent and requires an improvement of the state-of-the-art by 2-3 orders of magnitude.

## II. Ultrasensitive FIR Bolometers

Several approaches have been pursued in order to demonstrate an NEP $\sim 10^{-19}$-$10^{-20}$ W/Hz$^{1/2}$ in the FIR. They include kinetic inductance detectors [9-10], quantum dot detector [11] and bolometers with a superconducting transition-edge sensor (TES) thermometer. Because of the relatively simple underlying physics and long application heritage, the latter look more promising at the moment for meeting new sensitivity challenges.

Thermal energy fluctuation (TEF) (aka phonon) noise is dominating in a well-optimized bolometer and the corresponding NEP contribution is:

$$\text{NEP}_{\text{TEF}} = (4k_BT^2G)^{1/2}, \quad (1)$$

where G is the effective thermal conductance. G is commonly used as a benchmark when different bolometric devices are compared. A more traditional version of the TES bolometer uses a suspension made from thin and narrow Si$_3$N$_4$ to thermally isolate the radiation absorber and the TES thermometer from the heat sink. Here an impressively low thermal conductance has been achieved in geometrically isolated structures in [12]. At 65 mK (probably the lowest critical temperature one may expect for a practical space TES detector), G ≈ 30 fW/K was measured that corresponds to NEP$_{\text{TEF}}$ = $8\times10^{-20}$ W/Hz$^{1/2}$. Similarly low NEP$_{\text{TEF}}$ were obtained in the following works from the same group [13-14] where SiN mesh structures optimized for better radiation absorption and smaller time constant were used.

Another group [15] has achieved a fully functioning bolometer using a MoAu TES and a Ta radiation absorber suspended with long and narrow SiN beams. The TEF noise NEP has been reported to be $1.3\times10^{-18}$ W/Hz$^{1/2}$ at ≈ 100 mK. Similar detectors have been tested optically yielding the total NEP = $2\times10^{-18}$ W/Hz$^{1/2}$ at $\lambda$ ≈ 30-60 µm [15-16].

Our approach is to use the electron-phonon decoupling mechanism in a small TES in order to achieve an ultralow NEP [17]. In this case, a SiN membrane is not needed and the TES is fabricated directly on Si or sapphire substrate. When a TES is made from a thin (<100 nm) superconducting film, the temperature of electrons becomes greater than the phonon temperature when the radiation or current heating is applied. The phonon temperature remains close to that of the substrate since non-equilibrium thermal phonons escape from the film very fast. The characteristic time in this case is the electron-phonon energy relaxation time $\tau_{\text{e-ph}}$, which depends on the temperature but does not depend on the device volume. In turn, the effective thermal conductance $G_{\text{e-ph}} = C_e/\tau_{\text{e-ph}}$ ($C_e$ is the electron heat capacity) is proportional to the volume, so NEP$_{\text{TEF}}$ decreases in smaller devices. The practically useful device size is ~ µm or less so the device contacts should be made from a superconducting material with large $T_C$ in order in employ the Andreev reflection mechanism confining the electron thermal energy inside the device volume.

We have been using Ti hot-electron TES's in our work though some other materials (e.g., Hf, Ir, W) might be suitable too. Although the electron-phonon interaction strength varies to some degree between these materials the most important consideration is the availability of the fabrication technique leading to small devices with low $T_C$ and large enough sheet resistance $R_s$ = 20-50 Ohm. The latter is important for a good rf impedance match between the device and a microantenna or a waveguide which are the only ways to couple a subwavelength-size nano-HEB device to FIR radiation.

In our recent work [18], we pushed the device size to submicron (device volume of about $10^{-21}$-$10^{-20}$ m$^3$) and measured an extremely low thermal conductance $G_{e\text{-}ph}$. In comparison to the phonon conductance in SiN, the electron-phonon conductance in a metal has stronger temperature dependence. In fact, we achieve $G_{e\text{-}ph}$ = 0.3 fW/K at 65 mK that corresponds to NEP$_{TEF}$ = $9\times10^{-21}$ W/Hz$^{1/2}$. However, the very small TES devices studied in [18] did not have large enough critical current so they were not very useful for the noise studies or optical measurements. In the following works [19-20] somewhat larger size devices (6μm×0.4μm×56nm and 2μm×0.15μm×60nm) were used where a sufficiently large critical current along with $T_C$ = 330-350 mK was achieved due to an increased film thickness. Electrical noise and $G_{e\text{-}ph}$ have been studied as functions of temperature yielding an electrical NEP ≈ $2\times10^{-20}$ W/Hz$^{1/2}$ at 60 mK along with $\tau_{e\text{-}ph}$ ≈ 0.5 ms.

Although investigation of the electrical and thermal characteristics of bolometers is an important development step, the ultimate goal is optical demonstration of the detector sensitivity. In this work, we present the first results on the optical NEP measurements in Ti nano-HEBs coupled to planar antennas using an adequately low-power (femtowatt) radiation source. Besides the absolute NEP figures, an interesting feature in the detector response manifesting the presence of the photon shot noise was observed. The ability of the bolometer to detect photon noise is an independent confirmation of the validity of the optical power calibration scale in our setup.

### III. DEVICES AND EXPERIMENTAL SETUP

*A. Devices*

Our current device fabrication process is different from what was used in [18]. For this work, the nano-HEB devices were fabricated on c-axis sapphire using magnetron sputtering of Ti target and were patterned using optical lithography and ion beam milling. A 650 GHz twin-slot planar antenna structure was formed afterwards by means of the lift-off process using an NbTiN/Au bi-layer. This type of antenna with coplanar waveguide (CPW) impedance tuning sections has been well characterized and used by many groups since our original work [19]. NbN HEB mixers on the Herschel's HIFI use similar antennas but at THz frequencies. The critical temperature in NbTiN was greater than 10 K even in the presence of Au so it served as the Andreev reflection contact material as well. The device parameters are listed in Table I. $R_N$ in the Table is the device normal resistance.

Figure 1 shows a close-up of Ti device #1 in the center of the CPW structure. The real part of the antenna impedance derived from the HFSS modelling was ~ 30-40 Ohm at the resonance frequency of 650-670 GHz depending on the gap size between the CPW lines in the middle of the structure. Therefore, the impedance match should be very good for the device normal resistance values in Table I. The choice of the central frequency was driven by the trade-off between the intention to use the highest possible frequency and the size of the gaps between the central line in the CPW and the ground plane achievable using available optical lithographic equipment. The use of high frequency was important in order to achieve a better control over the radiation power emitted by a cryogenic black body calibration source (see next subsection). Beside the devices with twin-slot antennas many spiral antenna-coupled devices were fabricated on the same wafer, which have not yet been tested.

TABLE I
PARAMETERS OF Ti NANO-HEBS

| Device | Substrate | Length × Width (μm × μm) | Thickness (nm) | $T_C$ (mK) | $R_N$ (Ω) |
|---|---|---|---|---|---|
| 1 | sapphire | 2.0 × 1.0 | 20 | 357 | 45 |
| 2 | sapphire | 1.0 × 1.0 | 20 | 360 | 28 |

That the normal metal resistivity ρ = 45-55 μOhm cm of our films was similar to that of [20] for same film thickness. The $T_C$ in found in [20] was, however, much smaller (~ 100 mK). Transport and superconducting properties of our large area films and small devices will be addressed in a separate work in the future.

*B. The Setup*

All the measurements were performed in a dilution refrigerator. A dc SQUID was used as the readout. The devices were biased via a 1-Ω resistor connected in series with the device and with the SQUID input coil. The resistor was situated in the vicinity of the device on the mixing chamber. The SQUID amplifier was placed at the 1-K pot of the dilution refrigerator and was connected to the device circuit via a magnetically shielded superconducting twisted pair. The device assembly was in an rf tight superconducting shield in order to avoid overheating of the device by uncontrolled rf interferences and noise. The bias lines and the SQUID wires were fed through custom low-pass filters (LPF), which were built from discrete element LPFs placed inside metal tubes filled with lossy microwave absorbing compound. The SQUID noise (~ 2 pA/Hz$^{1/2}$) and bandwidth (~ 70 kHz) were calibrated in a separate experiment using the Johnson noise of a 1-Ω resistor at different temperatures.

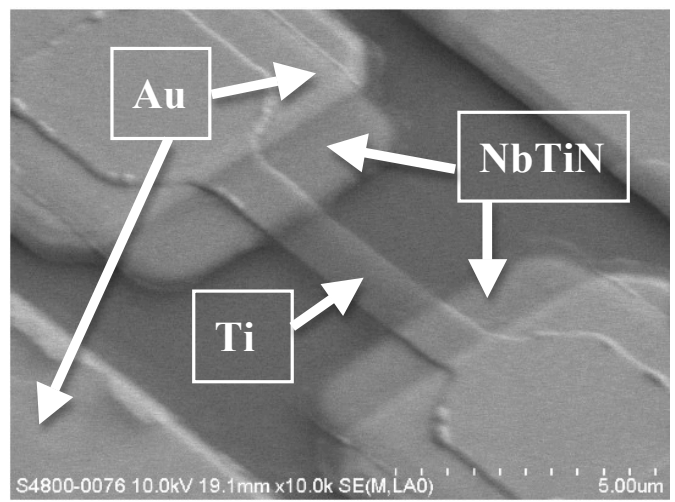

Fig.1. SEM image of a 2μm×1μm Ti nano-HEB device (device #1 from Table I) in the center of the CPW structure of the planar antenna.

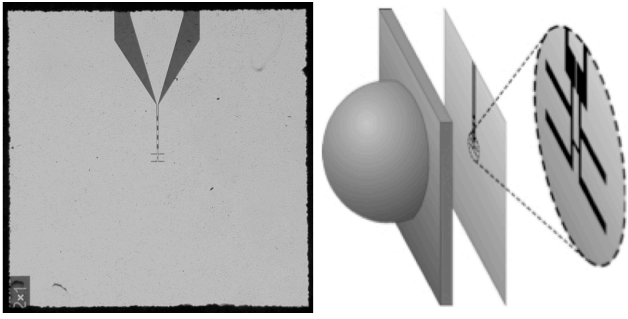

Fig.2. Detector chip layout and its attachment to the lens. The chip size is 5 mm × 5 mm. The lens diameter is 12 mm and is out of scale on this figure.

In order to perform measurements at different temperatures below $T_C$ a small superconducting solenoid with an iron core was mounted near the bolometer. The perpendicular field created by the solenoid was enough to suppress $T_C$ down to 50 mK with the solenoid current not exceeding 100 mA. The entire experimental volume was surrounded by a Cryoperm-10® magnetic shield immersed in liquid He4.

Optical NEP was measured using a black body source made from a metal disc in which a thermometer and a small heater were embedded. The emitting surface was machined into an array of small pyramids and was painted with an FIR absorbing paint whose high (97%) absorptivity was validated by both direct measurements of the reflection at around 600 GHz and comparison with the emissivity of foamed Eccosorb used as the standard calibration target for a heterodyne receiver. This black body assembly was weakly coupled to the 1K pot of the dilution refrigerator so its temperature $T_{BB}$ could be controlled with a better than 0.1 K precision from 1.5 K to 10 K.

The detector chip was glued to the backside of a 12-mm diameter lens made from the high-purity high-resistivity Si (ρ > 10 kOhm cm) thus forming a hybrid antenna (Fig. 2) with a well-defined and narrow diffraction limited beam. In order to define the optical bandwidth a bandpass (10% fractional bandwidth) mesh filter centred at 650 GHz was used [21]. The filter transmission spectrum Tr(ν) was characterized prior to the experiment using an FTS. We assumed that our hybrid antenna couples only into a single radiation mode and therefore the amount of radiation power $P_{rad}$ incident on the detector can be calculated as follows:

$$P_{rad} = \int_0^\infty \frac{Tr(\nu)h\nu d\nu}{\exp(h\nu/k_B T_{BB}) - 1}. \qquad (2)$$

In practice, a black body temperature range $T_{BB}$ =1.5-5 K was sufficient to observe the entire evolution of the detector output signal and of the noise up to the full saturation of the output (R ≈ $R_N$).

IV. EXPERIMENTAL RESULTS

*A. Procedures*

The experimental procedure for determination of NEP consisted of the measurements of bias current I as function of $T_{BB}$ varied in small 0.25 K steps and of the output system noise $i_n$. The corresponding optical power was calculated using Eq. 2. From the initial part of the I($P_{rad}$) dependence, detector responsivity $S_I = \Delta I/P_{rad}$ was calculated. Then the small signal optical NEP was found as $i_n(P_{rad}=0)/S_I$.

We also measured IV characteristics of the devices as function of temperature and radiation power. From the IVs taken and different temperatures, the effective thermal conductance corresponding to the device electron temperature $T_e$ can be extracted. This procedure is based on the assumption that the device resistance is a single-valued function of the electron temperature R($T_e$). Then from the

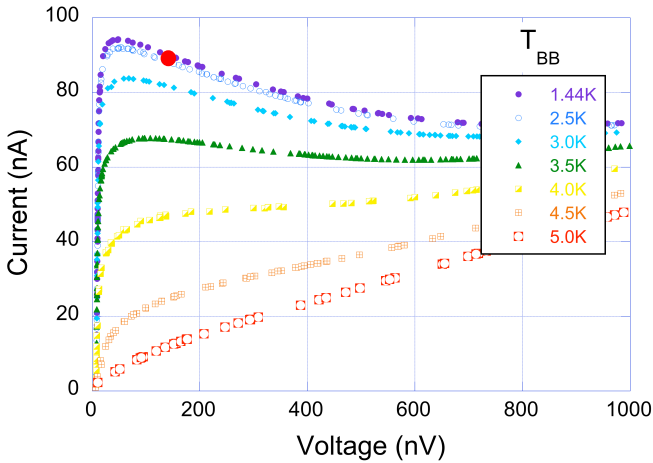

Fig. 3. IV characteristics at 355 mK (no magnetic field) under optical loading from the blackbody. $T_{BB}$ = 5 K corresponded to ≈ 160 fW of the radiation power emitted into a single mode through the filter bandwidth. The red circle denotes the operational point where the device was biased in the absence of radiation power.

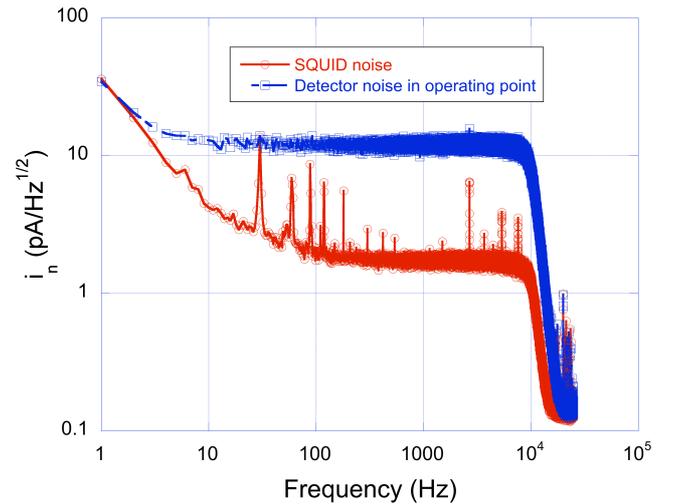

Fig. 4. Output noise at the bias point of Fig. 3 at 355 mK and SQUID noise. The latter was measured when the device was in the normal state (large resistance, small Johnson noise). A sharp cut-off at 10 kHz is due to the external bandwidth limiting filter.

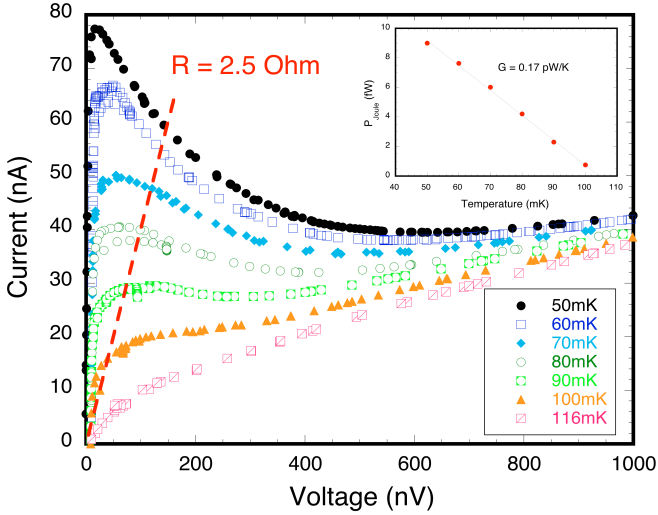

Fig. 5. IV characteristics in the 50-116 mK temperature range used for determination of the thermal conductance (see inset). The bias point was at the crossing of the 50-mK IV curve at the 2.5-Ohm resistance line.

heat balance equation one can obtain:

$$(P_{Joule}1 - P_{Joule}2) = G(T2-T1), \quad (3)$$

where $P_{Joule}$ is the Joule power dissipated in a given bias point. Thermal conductance G derived in this way was used to calculate $NEP_{TEF}$. Equation 3 assumes that T2-T1 << T1, T2 but can be easily generalized for a non-linear case.

In a similar fashion, the optical coupling efficiency η can be derived from the IVs taken at a fixed bath temperature but when $P_{rad}$ is varied. In this case, the following equation holds:

$$P_{Joule}1 - P_{Joule}2 = \eta(P_{rad}2 - P_{rad}1). \quad (4)$$

The coupling efficiency derived from Eq. 4 was crosschecked with the ratio $NEP_{TEF}/NEP$, which should yield the same η value.

*B. The Data*

Here we present the detailed data only for device #2. More details for device # 1 can be found in [23]. Figure 3 demonstrates optical loading of the device by the 650 GHz radiation. From these IVs, an optical coupling efficiency η = 40-80 % was estimated. The large error margin is due to the fact that η happened to depend on the particular R = constant line along which the data were taken (see Eq. 4).

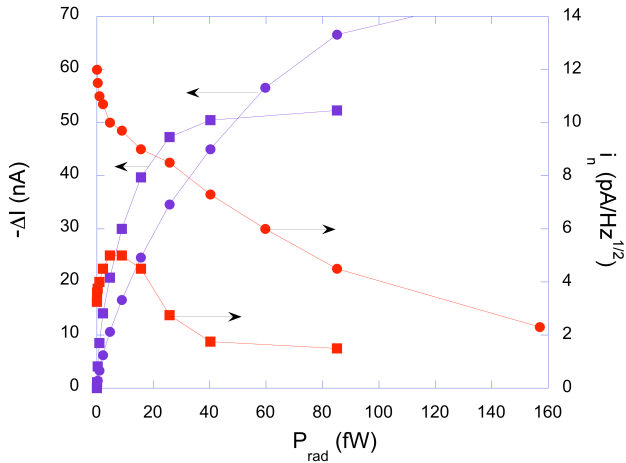

Fig. 6. Experimental data used for determination of optical NEP. Round symbols: T = 355 mK; square symbols: T = 50 mK.

Figure 4 demonstrates the detector output noise at the operating point of Fig. 3. The noise spectrum is very flat in this 10-kHz bandwidth. A 1/f-noise below a few Hz originates in the SQUID. Its origin is still unknown.

Figure 5 shows IV characteristics when a certain magnetic field was applied. From these data, the thermal conductance was obtained. $P_{Joule}$ vs T dependence in the inset in Fig. 5 shows that the electron temperature for this resistance was $T_e$ = 105 mK. The broad transition in magnetic field was likely the reason for the output noise at 50 mK to be much smaller than that at 355 mK (see Fig. 6). Since $T_e$ is substantially greater than T = 50 mK a slightly different expression from Eq. 1 should be used for determination of $NEP_{TEF}$:

$$NEP_{TEF} \approx (2k_B T_e^2 G)^{1/2}. \quad (5)$$

Equation 5 is valid in a strong non-equilibrium case reflecting the fact that the contribution of the fluctuation of energy at the bath temperature is negligible compared to the fluctuation of energy at the electron temperature.

For T=355 mK, $NEP_{TEF}$ was determined in similar fashion, i.e., using IVs at different temperatures. However, the temperature range where IVs differ from each other was much more narrow ($T_e$ = 367 mK). This is because the natural superconducting transition in the absence of the magnetic field is relatively narrow (within 10 mK). In this case the difference between T and $T_e$ is not so significant so the fluctuations at both temperatures contribute to $NEP_{TEF}$.

Finally, optical NEP was determined at both 50 mK and 355 mK by measuring the change of current vs radiation power and the output noise at the same time (see Fig. 6). We found an order of magnitude difference in the NEP at these two temperatures (see Table II). Saturation of ΔI vs $P_{rad}$ naturally occurs much sooner at 50 mK than at 355 mK. The output noise also behaves differently. Whereas the noise at 355 mK monotonically decreases with $P_{rad}$, the noise at 50 mK exhibits a peak at some power range below 20 fW. We speculate that the origin of this peak may be in the detection of fluctuation of power in the impinging radiation. This photon shot noise is characterized by

$$NEP_{phot} = (2P_{rad}h\nu)^{1/2}. \quad (6)$$

When the detector becomes sensitive enough to detect this fluctuation (NEP ~ $NEP_{phot}$) the output noise-like signal increases as square root of $P_{rad}$. Eventually, $S_I$ drops and the output noise decreases. This effect could not be seen at 355 mK since a much greater $P_{rad}$ ~ 100 fW is needed to make $NEP_{phot}$ large enough to be detected. But at such high power the detector is already saturated.

Another argument in favor of this explanation is the comparison between the $P_{rad}$ scales for the devices with different sensitivity. Figure 7 shows the output noise vs $P_{rad}$ for both devices. Device #1 was operated at 100 mK but the electron temperature was about 150 mK [23]. Also device #1

   4

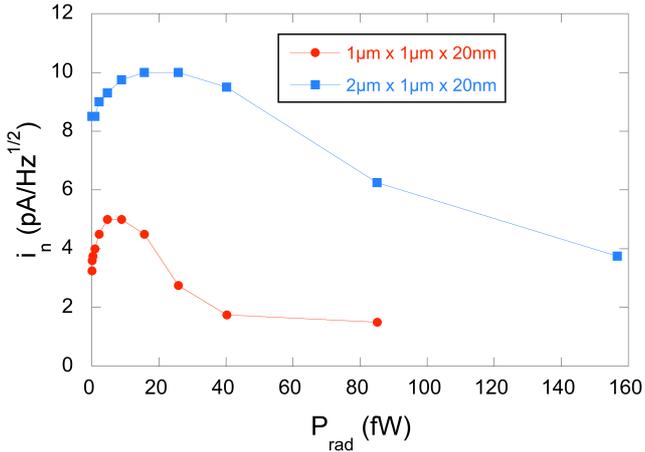

Fig. 7. Output noise for both devices #1 and #2 in magnetic field at temperatures well below $T_C$.

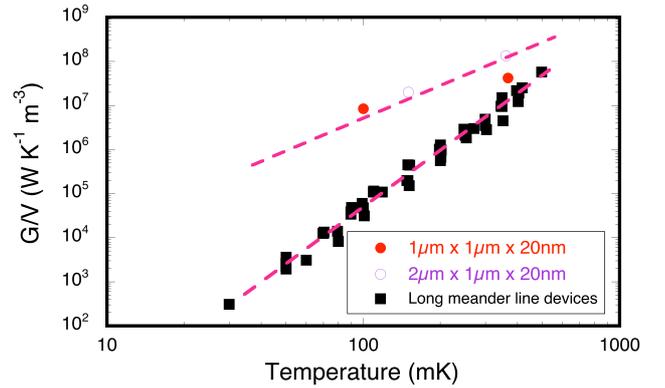

Fig. 8. Thermal conductance in large and small hot-electron bolometers normalized to the device volume. Round symbols are the data for the devices of Table I. Long meander data are from [-].

is at least 2 times larger in volume than device #2. As a result, it had optical NEP = $1.4 \times 10^{-18}$ W/Hz$^{1/2}$, that is, a factor of 4-5 greater than the NEP in device #2. One can see from Fig. 7 that the position of the noise peak and the radiation power required to saturate the detector was correspondingly greater for device #1

## V. DISCUSSION

Table II summarizes the main results of this study. The NEP data look very encouraging especially for device #2, which is already meeting the sensitivity requirements for SAFARI on SPICA. This NEP is an order of magnitude smaller than the closest competing TES approach using SiN membranes [15-16] has demonstrated. The optical coupling efficiency found from the ratio of NEP$_{TEF}$ and NEP correlates with η derived from the shift of IV curves under optical loading. To be fair, the η value may be slightly overestimated given the fact that the Si lens did not have any AR coating. This is, however, quite understandable: these are very first measurements and more work is needed to tweak the measurement procedures. We are planning on using an additional high-pass 600 GHz filter with a sharp cut-off in order to eliminate any possible leak of the out-of-band radiation power at the long-wavelength wing of the mesh filter. Also, devices with spiral antennas will be tested to study the effect of different antenna types. Such devices will be eventually needed for FIR applications of this detector (λ < 100 μm) where slot antennas can be difficult to fabricate.

TABLE II

NEP DATA

| Device | $T_e$ (mK) | NEP (aW/Hz$^{1/2}$) | NEP$_{TEF}$ (aW/Hz$^{1/2}$) | NEP$_{TEF}$/NEP |
|---|---|---|---|---|
| 1 | 150 | 1.4 | 1.0 | 0.71 |
|   | 357 | 8.6 | 6.3 | 0.73 |
| 2 | 105 | 0.30 | 0.23 | 0.77 |
|   | 367 | 3.0 | 2.5 | 0.83 |

Another important issue is a comparison of the achieved NEP with the expectations for this type of detectors. Since the optical NEP came close to NEP$_{TEF}$, the latter can be used for comparison since it is easy to estimate from the material parameters. In our original paper [17] we predicted that NEP in 1μm × 1μm size nanobolometers at 100 mK should be at least ~$10^{-20}$ W/Hz$^{1/2}$. The best current NEP$_{TEF}$ is an order of magnitude worse. Comparison with the large area devices gives some clue on what might be wrong with the small devices. Figure 8 shows a normalized by volume electron-phonon thermal conductance measured in 4 long-meander-line devices (some up to 10 cm long) over the years [24-25]. They were all fabricated on sapphire using either e-beam evaporation or magnetron sputtering. Amazingly, all the data points closely follow one universal line that is a strong indication of the electron-phonon nature of G in these devices, which should be proportional to the volume. When the G-data for devices #1 and #2 are plotted on the same graph it becomes apparent that there is an excessive thermal conductance in these device especially at low temperature where magnetic field was used. The excessive thermal conductance is 100-time greater than should be due to the electron-phonon interaction alone. We suspect that the reason for that might be in the unfavorable configuration of metal layers in the device contact areas. As seen in Fig. 1, the projection of top Au layer overlaps with Ti being separated by NbTiN. When a perpendicular magnetic field is applied it suppresses $T_C$ in Ti. The field is not sufficient to destroy superconductivity in NbTiN which has $T_C$ > 10 K. However, the field may create a lot of magnetic vortices in the NbTiN layer. These vortices have normal metal core, which would connect the Ti layer with the Au layer thus creating a channel for cooling of hot electrons via electron diffusion.

The next iteration of the devices will not have the Au layer which is redundant for the purpose of low electromagnetic loss. NbTiN by itself should be a good enough rf conductor at 650 GHz.

This quick fix may help to get even better NEP data relatively soon. However, in a long run, one need to get rid of the magnetic field which is unacceptable for the detectors operating in large (1000s pixels) arrays. We see several options for reducing $T_C$ to about 100 mK. The simplest approach is to keep reducing the films thickness. However the transition region between "high" (~ 300 mK) and "low" (~ 100 mK) $T_C$ may be narrow so a large scattering of device-to-device parameters can happen. Another approach is implantation of $^{55}$Mn$^+$ ions into ready devices or into bare Ti films. This has been demonstrated be a reasonably well controlled process [26] and we plan to try it within next few months with devices similar to those of Table I. A similar techniques has been also used successfully for tuning TC in W using $^{56}Fe^+$ ions [27]. One more option is to synthesize some binary alloys (like, e.g. AlMn [28]), which will require a significant material development effort. We should mention that bi-layer materials commonly used for membrane supported TES will not work here because of their low resistivity making it impossible to match the device $R_N$ to the antenna impedance.

Finally, some SQUID based multiplexing technique should be made compatible with the hot-electron nano-HEBs. There may multiple solutions here depending on what sensitivity and, correspondingly, what operating temperature is targeted. For the most sensitive applications when T = 50-100 mK, the time constant in nano-HEB devices would be ~ 0.5-1 ms [23]. If it turns out that the thermal conductance in these devices will be eventually reduced to the value of large area Ti films then $\tau_{e-ph}$ = 0.6 ms at 100 mK and $\tau_{e-ph}$ = 2.5 ms at 65 mK. This is derived using the Sommerfeld constant value $\gamma$ = 315 J m$^{-3}$ K$^{-2}$ and the data of Fig. 8. There need to be some design study in order to understand if the well-established time-domain multiplexing (TDM) technique can be used. However, the recently demonstrated GHz frequency domain multiplexing (FDM) using microwave SQUIDs [29-30] should work well not only at 100 mK but also at 300-400 mK where $\tau_{e-ph}$ is in the μs range. This technique is still emerging but should soon become suitable for reading large arrays.

## VI. CONCLUSIONS

In conclusion, a significant progress has been made in the development of the nano-HEBs based detector technology resulting in the first optical demonstration of the state-of-the-art detector sensitivity in FIR. The current generation of the detectors together with the microwave SQUID readout is suitable for building a detector array with $NEP \sim 10^{-17}$ W/Hz$^{1/2}$ operating in a He3 dewar. Such systems can be used for photometry, polarimetry, or CMB studies from balloons or SOFIA airplane.

The obtained record low NEP = $3\times10^{-19}$ W/Hz$^{1/2}$ is a good news for the low-background spectroscopy in space where no adequate detectors have existed so far. This already meets the sensitivity goals for the SPICA/SAFARI instrument. The future R&D work will address the ways of getting optical NEP down to $10^{-20}$ W/Hz$^{1/2}$ by means of controllable $T_C$ reduction in Ti nano-HEB devices and also by making the devices smaller (submicron). A significant improvement of the existing optical setup will be required in order to control the fW level optical power with sufficient precision.

## ACKNOWLEDGMENT

This research was carried out at the Jet Propulsion Laboratory, California Institute of Technology, under a contract with the National Aeronautics and Space Administration. The authors are grateful to D. Harding and S. Pereverzev for the help with optical NEP setup and cryogenic filters, to N. Llombart for the help with HFSS modelling, and to J. Kawamura for many fruitful discussions and for providing a superconducting solenoid.